%% file: main.tex
\documentclass[sigconf,natbib=true]{acmart}
\usepackage{booktabs}
\usepackage{enumitem}
\usepackage{multicol}
\usepackage{multirow}
\usepackage{threeparttable}
\usepackage[most]{tcolorbox}
\usepackage{bbding}
\usepackage{amsmath}
\usepackage{todonotes}
\usepackage{pifont}

\AtBeginDocument{%
  }

\copyrightyear{2025}
\acmYear{2025}
\setcopyright{cc}
\setcctype{by}
\acmConference[SIGIR-AP 2025]{Proceedings of the 2025 Annual International ACM SIGIR Conference on Research and Development in Information Retrieval in the Asia Pacific Region}{December 7--10, 2025}{Xi'an, China}
\acmBooktitle{Proceedings of the 2025 Annual International ACM SIGIR Conference on Research and Development in Information Retrieval in the Asia Pacific Region (SIGIR-AP 2025), December 7--10, 2025, Xi'an, China}
\acmDOI{10.1145/3767695.3769490}
\acmISBN{979-8-4007-2218-9/2025/12}

\begin{document}



\input{Headings/head}
\input{Headings/abstract}
\input{Headings/keywords}
\maketitle
\input{Sections/1}
\input{Sections/2}
\input{Sections/3}
\input{Sections/4}
\input{Sections/5}



\bibliographystyle{ACM-Reference-Format}
\balance
\bibliography{main}

\clearpage
\appendix

\end{document}

%% file: Headings/head.tex
%
%

\title{Human vs. Agent in Task-Oriented Conversations}

\author{Zhefan Wang}
\affiliation{%
  \institution{DCST, Tsinghua University}
  \city{Beijing}
  \country{China}}
\email{wzf23@mails.tsinghua.edu.cn}

\author{Ning Geng}
\affiliation{%
  \institution{Emory University}
  \city{Atlanta}
  \state{Georgia}
  \country{USA}}
\email{ngeng3@emory.edu}

\author{Zhiqiang Guo}
\affiliation{%
  \institution{DCST, Tsinghua University}
  \city{Beijing}
  \country{China}}
\email{georgeguo.gzq.cn@gmail.com}

\author{Weizhi Ma}
\authornote{Corresponding author. This work is supported by the Natural Science Foundation of China (Grant No. U21B2026, 62372260) and the National Key Research and Development Program of China under Grant 2024YFC3307403.}
\affiliation{%
  \institution{AIR, Tsinghua University}
  \city{Beijing}
  \country{China}}
\email{mawz@tsinghua.edu.cn}

\author{Min Zhang}
\authornotemark[1]
\affiliation{%
  \institution{DCST, Tsinghua University}
  \city{Beijing}
  \country{China}}
\email{z-m@tsinghua.edu.cn}

\renewcommand{\shortauthors}{Zhefan Wang, Ning Geng, Zhiqiang Guo, Weizhi Ma, and Min Zhang}



%% file: Headings/abstract.tex
\begin{abstract}
Task-oriented conversational systems are essential for efficiently addressing diverse user needs, yet their development requires substantial amounts of high-quality conversational data that is challenging and costly to obtain.
While large language models (LLMs) have demonstrated potential in generating synthetic conversations, the extent to which these agent-generated interactions can effectively substitute real human conversations remains unclear.
This work presents the first systematic comparison between LLM-simulated users and human users in personalized task-oriented conversations.
We propose a comprehensive analytical framework encompassing three key aspects (conversation strategy, interaction style, and conversation evaluation) and ten distinct dimensions for evaluating user behaviors, and collect parallel conversational datasets from both human users and LLM agent users across four representative scenarios under identical conditions.

Our analysis reveals significant behavioral differences between the two user types in problem-solving approaches, question broadness, user engagement, context dependency, feedback polarity and promise, language style, and hallucination awareness.
We found consistency in the agent users and human users across the depth-first or breadth-first dimensions, as well as the usefulness dimensions.
These findings provide critical insights for advancing LLM-based user simulation. 
Our multi-dimensional taxonomy constructed a generalizable framework for analyzing user behavior patterns, offering insights from LLM agent users and human users.
By this work, we provide perspectives on rethinking how to use user simulation in conversational systems in the future.\footnote{Code and data are available at \url{https://github.com/wzf2000/RecLLMSim/tree/Human_Vs_Agent}.}
\end{abstract}

%% file: Headings/keywords.tex
\keywords{Large Language Models; User Simulation; Task-Oriented Conversational Systems}

\ccsdesc[500]{Information systems~Information retrieval}

%% file: Sections/1.tex
\section{Introduction}





The rapid advancement of large language models (LLMs) has catalyzed significant interest in leveraging their capabilities for user simulation.
Such simulations offer compelling advantages: they can efficiently generate vast quantities of human-like interaction data, thereby accelerating the development and iteration cycle of conversational systems, such as recommender agents.
Furthermore, simulated users present a viable solution for privacy preservation by reducing reliance on sensitive real-user data, while also serving as a cost-effective alternative to expensive A/B testing.
By providing robust offline evaluation environments, LLM-based simulations enhance the credibility of experimental results and significantly lower the costs associated with online validation of new algorithms or system optimizations.
Additionally, the ability to simulate diverse user personas with specific demographic or behavioral traits enables broader exploration and enhancement of personalized system interactions.

However, the effectiveness of simulation-driven approaches critically depends on the fidelity and reliability of the underlying LLM-simulated users.
Researchers are increasingly employing LLM-based user simulations, yet the fundamental question of how closely these simulations replicate the behavioral characteristics of real human users remains inadequately addressed.
While prior research~\cite{zhu2024reliable,zhu2025llm} has identified various limitations in user simulations—such as unnatural conversational patterns, limited contextual depth, or inconsistent personal adherence, particularly noted in conversational recommender systems (CRS)—a significant gap remains.
Systematic, direct comparisons of LLM-simulated user behaviors against actual human user behaviors under rigorously controlled, near-identical conditions are notably scarce.
Understanding the precise nature and magnitude of behavioral discrepancies is crucial for enhancing the realism and utility of simulations.

To address this gap, we conducted a controlled experiment collecting parallel interaction datasets from both human participants and LLM-simulated users within an almost identical conversational setting.
Building upon detailed observations of these interactions, we performed a comparative behavioral analysis, focusing on key predefined interaction dimensions, including all-in-one or step-by-step, question broadness, breadth-first or depth-first, user engagement, context dependency, feedback polarity and explanation, language style, user satisfaction, hallucination awareness, and usefulness.
Utilizing automated classification and scoring methodologies, we quantitatively and qualitatively contrasted the conversational behaviors across the two user groups.

Our analysis reveals consistency and significant differences between LLM-simulated user behaviors and their human counterparts.
These differences manifest in specific, measurable aspects of interaction that impact the realism and effectiveness of simulations for downstream tasks.
Based on these empirical findings, we distill concrete actionable insights and present a set of targeted recommendations for refining LLM-based user simulation frameworks.
Our work lays the foundation for developing more accurate, human-aligned simulations, ultimately enhancing their value in developing and evaluating interactive AI systems.

In summary, the main contributions of this work can be summarized as follows:
\begin{itemize}[leftmargin=*, parsep=2pt]
    \item This study presents the first systematic comparison between human users and LLM agent users in task-oriented conversations, addressing a critical gap in conversational AI research.
    \item We propose a structured comparison framework encompassing three key aspects and ten detailed dimensions, enabling a fine-grained evaluation of user differences.
    \item Our findings reveal distinct and consistent characteristics of LLM agent users compared to humans, providing fundamental insights to aid future LLM-based user simulation.
\end{itemize}

%% file: Sections/2.tex
\begin{figure*}[ht]
    \includegraphics[width=0.7\textwidth]{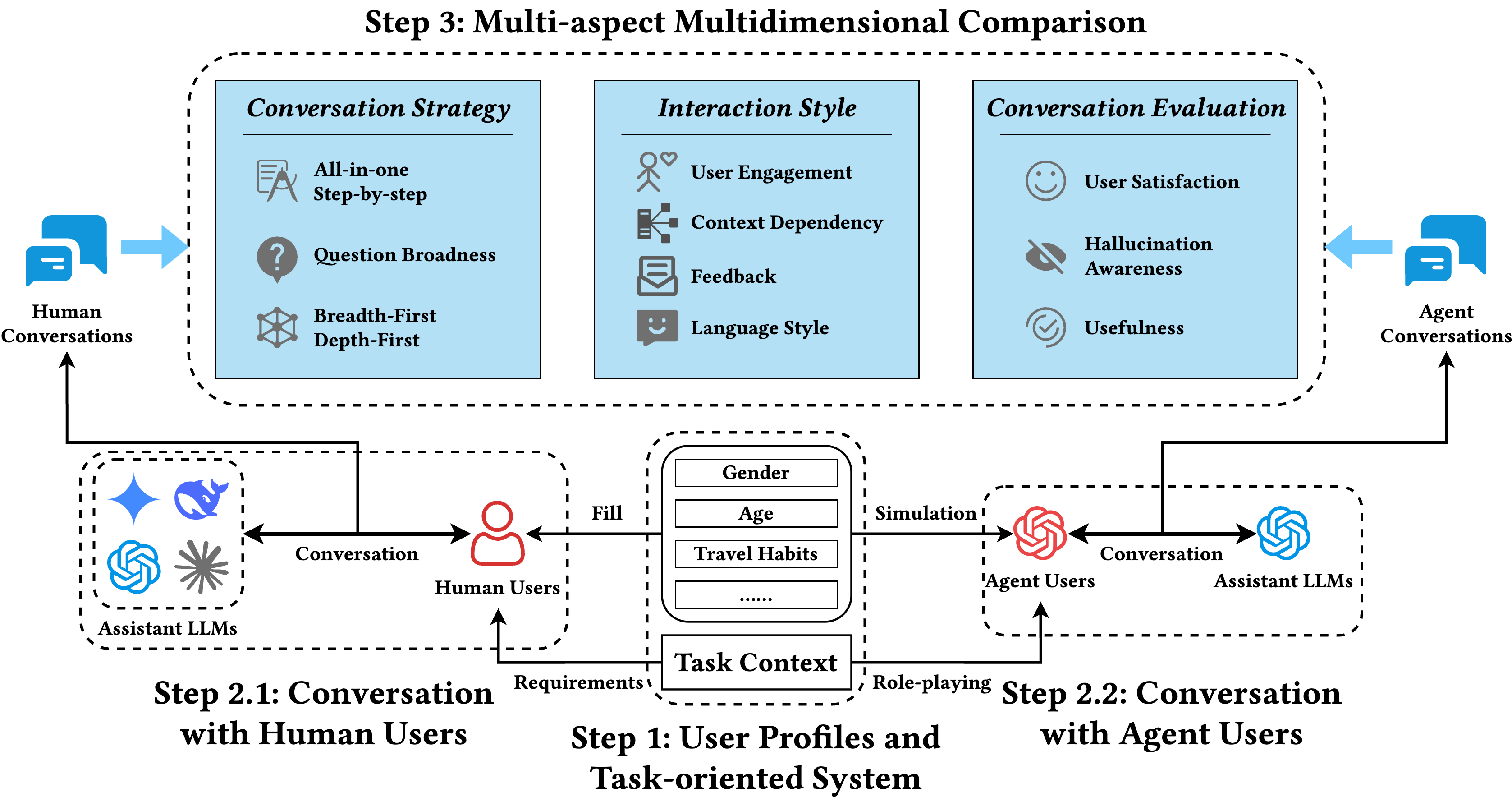}
    \caption{The overview of our study, including three main steps: user profiles and task-oriented system, conversation with human users or agent users, and multi-aspect multidimensional comparison.}
    \Description{}
    \label{fig:framework}
\end{figure*}

\section{Related Work}

\subsection{User Simulation}

User simulation has emerged as a valuable technique across various research domains, enabling cost-effective testing and evaluation without requiring extensive data from real users.
Before LLMs, user simulators were widely employed for evaluating recommender systems~\cite{afzali2023usersimcrs,zhang2020evaluating} and task-oriented conversational systems~\cite{sun2021simulating,sekulic2022evaluating}.

The advent of LLMs has significantly expanded simulation capabilities, revolutionizing multiple computational domains.
These models excel in data augmentation by generating high-quality synthetic data~\cite{askari2023generating,huang2023chatgpt,whitehouse2023llm} and enable sophisticated social behavior modeling through multi-agent simulations~\cite{park2023generative}.
Additionally, LLMs have proven valuable for automating conversational system evaluation~\cite{chan2023chateval,hu2023unlocking} and enhancing recommender systems through agent-based approaches~\cite{wang2023recagent,zhang2024agentcf,wang2023rethinking,zhu2024reliable}.
These diverse applications demonstrate the versatility of LLM agents in addressing complex simulation challenges across various fields.
Following these studies, we adopt the agent-based user simulation for our analysis.

The fidelity of user simulation is crucial for ensuring that insights derived from simulated interactions can be effectively transferred to real-world scenarios.
While numerous studies employ user simulators as proxies for real users, direct behavioral comparisons between simulated and human users in interactive settings remain scarce.
Potential differences between simulated users and real human users may lead to non-negligible bias in user simulation.
Our work addresses this gap by systematically comparing the conversational behaviors and interaction patterns of LLM-simulated users with those of real human users in task-oriented conversation contexts.

\subsection{Task-Oriented Conversational System}

Task-oriented conversational systems aim to assist users in accomplishing specific objectives through natural dialogue, addressing real-world needs such as booking services, finding information, or completing transactions~\cite{zhang2019dialogpt,peng2021soloist}.
These systems have become increasingly prevalent in practical applications, including customer service, travel planning, restaurant reservations, and technical support, reflecting the diverse task-completion scenarios users encounter in daily life.

The emergence of LLMs has transformed approaches to building task-oriented conversational systems.
While early work explored zero-shot applications of LLMs to these tasks~\cite{hudevcek2023llms}, subsequent research demonstrated that fine-tuning significantly improves performance~\cite{hosseini2020simple,gupta2022instructdial}.
More recent advances include instruction-tuning LLMs for specific subtasks, such as API argument filling~\cite{mok2024llm}, leveraging description-based approaches~\cite{zhao2022description}, and developing sophisticated simulation frameworks~\cite{luo2024duetsim,sekulic2024reliable}.
The automation of task-oriented chatbot development has also gained attention, with LLMs enabling more efficient creation of conversational flows~\cite{sanchez2024automating}.

However, current task-oriented conversational systems predominantly prioritize task completion metrics over user-centric considerations.
In contrast, our work explores more nuanced conversational scenarios that emphasize personalization and user profiles, where objectives are less clearly defined and conversations lack explicit termination signals.

\subsection{Previous Observations in Human vs. LLMs}


Existing comparative work primarily focuses on fields such as game theory, rationality, and decision-making tasks, while paying less attention to the comparison of direct behavior during conversations.

In game-theoretic settings, research~\cite{xie2024can} showed that LLM agents systematically differ from humans in trust-based interactions, typically exhibiting higher trust levels, while \citet{fontana2025nicer} discovered that LLMs behave more cooperatively than humans in the Prisoner's Dilemma.
\citet{alsagheer2024comparing} found that while LLMs show reasoning errors similar to humans, these similarities are superficial, and both groups respond differently to the same prompting strategies.
Most recently, researchers have revealed that although LLMs display similar aggregate-level biases to humans~\cite{horowitz2025llm}, these patterns emerge from fundamentally different processes.
LLMs exhibit strong recency biases and lack sophisticated phenomena, such as "surprise triggers change", which is observed in human behavior.

Under conversational settings, \citet{huijzer2023large} demonstrated that LLMs exhibit behaviors remarkably similar to humans in controlled experiments.
\citet{jiaqi2025comparative} highlighted that current human-LLM comparisons often rely on surface-level language similarities rather than deeper behavioral understanding.

While these studies provide valuable insights through controlled experiments and game-theoretic scenarios, they primarily focus on isolated decisions or simple strategic interactions.
Our work extends this research to personalized task-oriented conversations, examining behavioral differences in dynamic, multi-turn conversations where humans and LLM agent users pursue specific communicative goals through natural language interaction.

%% file: Sections/3.tex

\section{Conversation Protocol and Study Settings}

\subsection{Study Overview}

Our study was conducted in three main steps.
As shown in Figure~\ref{fig:framework}, we first developed a task-oriented conversational system with user profiles (Sections~\ref{Section-scenarios} and \ref{Section-profiling}).
Then, we conducted and collected conversations from two types of users under nearly identical conversational settings, using a standardized protocol (Sections~\ref {Section-assistant}, \ref{Section-simulation}, and \ref{Section-human}).
Finally, we employed human self-report, third-party human annotations, and LLM automated analysis to systematically compare user behaviors across multiple aspects and dimensions (Section~\ref{Section-settings}, \ref{Section-4}, \ref{Section-5}, and \ref{Section-6}).
This controlled approach ensured direct comparability between the two user types while maintaining ecological validity.

\subsection{Conversation Scenarios}
\label{Section-scenarios}


During the conversation collection process, we considered four distinct conversational scenarios: preparing gifts, travel planning, recipe planning, and skills learning planning.
These scenarios were selected to cover a broad spectrum of daily life situations requiring planning, encompassing domains such as clothing, food, housing, transportation, and leisure activities.

The specific configuration of task requirements has a significant impact on both the conversation process and the quality of outcomes across different conversational scenarios.
Therefore, for each scenario, we defined task requirements by incorporating multiple configurable variables alongside fixed parameters.
These included, for example, temporal constraints in travel planning (e.g., duration, deadlines), recipient-specific requirements in gift preparation (e.g., relationships, occasions), and scalability parameters in recipe planning (e.g., serving size, dietary restrictions).
Table~\ref{tab:scenario} shows the details of constraints for each scenario.

\begin{table}[htbp]
  \centering
  \caption{Each task description for the specified scenario is constructed from a prompt template with several constraints.}
  \begin{threeparttable}[c]
    \begin{tabular}{cc}
      \toprule
      Scenario & Fields \\
      \midrule
      Travel Planning & Time, Duration, Type, Destination  \\
      Recipe Planning & Reason, Scale, Level \\
      Preparing Gifts & Target, Reason \\
      Skills Learning Planning & Skill, Level, Reason \\
      \bottomrule
    \end{tabular}
  \end{threeparttable}
  \label{tab:scenario}
\end{table}


\subsection{User Profiling}
\label{Section-profiling}

Since user profiles directly influence conversation analysis, we aim to employ carefully controlled profiles during agent user simulations.
To achieve this, we utilized an LLM to generate a standardized set of user profiles in a unified batch process.
These profiles underwent manual filtering to establish the final pool of profiles.
Each profile within this pool comprises the following nine attributes: 1.\textbf{Gender} (e.g., Female); 2.\textbf{Age} (e.g., 55); 3.\textbf{Personality} (e.g., Compassionate, patient, nurturing); 4.\textbf{Occupation} (e.g., School Counselor); 5.\textbf{Daily Interests/Hobbies} (e.g., Reading psychology books, gardening, volunteering); 6.\textbf{Travel Habits} (e.g., Prefers staycations or short countryside trips); 7.\textbf{Dining Preferences} (e.g., Home-cooked meals, balanced diet); 8.\textbf{Spending Habits} (e.g., Saves for retirement, spends on family/grandchildren); 9.\textbf{Other Aspects} (e.g., Church attendance, book club membership).

Attributes 5–8 each correspond to a scenario (Section~\ref{Section-scenarios}) and represent scenario-aligned preferences (e.g., travel habits for travel planning), while attributes 1-4 reflect broader demographic and behavioral traits.


For the agent user, a profile was randomly assigned from this pool before each conversation simulation.
For the human user, profile selection was based on self-reported choices from the pre-established profile pool.
Specifically, participants were presented with multiple tags per attribute to select from, enabling a multi-tag representation of their profile.
Considering both the effectiveness and computational cost of LLMs, we selected GPT-4 to generate the initial set of the profile pool.

\subsection{Assistant LLMs}
\label{Section-assistant}

Throughout the conversation process, a large language model (LLM) served as the intelligent assistant to help users address task-related queries.
Before each conversation, the assistant LLM was provided with a basic scenario description, without access to specific task requirements or user profiles, ensuring a standardised starting point.

For human-user conversations, we randomized the assistant LLM across multiple high-performance API-based models, including the GPT series, Claude 3.7, Gemini 2.0, and Deepseek V3.
Human participants were not informed of the assigned LLM (i.e., the model remained anonymous to them).
For agent user conversations, we consistently employed the GPT-series as the assistant LLM to ensure uniformity in simulated interactions.

\subsection{User Simulation with LLM Agents}
\label{Section-simulation}



Before initiating conversations within a given scenario, LLM-based agent users were primed through multi-turn instructions.

The guided instructions were provided as contextual prompts before connecting the agent user with the assistant LLM.\footnote{Detail prompt can be found in \url{https://github.com/wzf2000/RecLLMSim/tree/Human_Vs_Agent/config/gen_chat_new.json}.}
After each assistant's response, the agent user was explicitly queried about conversation termination, enabling dynamic length control based on task completion status rather than fixed turns.

Through the above protocol, we obtained 1856 simulated conversations with 85 distinct user profiles.

\subsection{Human Users}
\label{Section-human}

We recruited human participants to engage in conversations under the aforementioned experimental setup.
The data collection protocol consisted of the following steps.

Before starting the conversation, each participant was required to complete their authentic user profile (consistent with the 9-attribute structure defined in Section~\ref{Section-profiling}) before conversation initiation.
Participants were then randomly assigned a task requirement from the target scenario and instructed to carefully review it before starting the conversation.

During the conversation process, participants were instructed to immerse themselves in the task context fully and strictly adhere to the given requirements throughout the conversation.
They maintained autonomy to terminate the conversation upon satisfactory task completion.
A safeguard mechanism allowed participants to request task replacement if the assigned requirement was deemed unrealistic or incompatible with their real-world experience.

During the experiment, a total of 146 participants were involved (81 male and 65 female).
Their professional backgrounds spanned nine distinct categories (e.g., Information Science \& Technology, Natural Sciences, and Engineering \& Applied Sciences).
We collected a total of 2,124 human-user conversations.
On average, each user engaged in 14.5 conversations, with an average participation time of between two and three hours per user.
While the majority of participants completed conversations across all four scenarios, a small number did not finish all scenarios due to time constraints or efficiency considerations.
Crucially, we ensured participation from at least 138 human users in each scenario.

\subsection{Empirical Study Settings}
\label{Section-settings}


\input{Tables/statistics}

Based on the collected conversation corpus (basic statistics shown in Table~\ref{tab:dataset_basic}), we conducted a comprehensive analysis of user behaviors and conversational characteristics through three primary aspects:
\begin{itemize}[leftmargin=*, parsep=2pt]
    \item \textbf{Conversation Strategy}: All-in-one or step-by-step, question broadness, and breadth-first or depth-first.
    \item \textbf{Interaction Style}: User engagement, context dependency, feedback polarity and promise, and language style.
    \item \textbf{Conversation Evaluation}: User satisfaction, hallucination awareness, and usefulness.
\end{itemize}

The analysis employed a triangulated assessment methodology combining quantitative, human-evaluated, and automated techniques.
Direct statistical measures extracted fundamental conversation features, while human perspectives were captured through dual channels: third-party annotations for agent-user conversations complemented by participant self-reports for human-user interactions.
For scalable, fine-grained analysis, we implemented a standardized LLM-based evaluation pipeline utilizing GPT-4o.
We utilized automatic evaluation on the following dimensions: all-in-one or step-by-step, question broadness, breadth-first or depth-first, context dependency, feedback polarity and promise, language style, and usefulness.
Human annotations evaluate user satisfaction and hallucination awareness.

Evaluation stability was enhanced by triplicate sampling with fixed model and input configurations.

\input{Tables/manual}

As mentioned, certain analytical dimensions employed automated LLM-based analysis, necessitating validation of their reliability.
To address this, we conducted further manual verification on a representative sample of the conversations (25 conversations per scenario, involving both user types).
Table~\ref{tab:manual_eval} showed that we achieved over 89\% inter-rater agreement across all dimensions, which demonstrated consistent reliability between automated and human analysis.

%% file: Tables/statistics.tex
\begin{table}[ht]
  \centering
  \caption{Statistics of the collected conversations from both users. "\#Tasks", "\#Users", and "\#Conv." represent the number of task requirements, users (with different profiles), and conversations, respectively.}
    \begin{tabular}{cc|ccccccc}
      \toprule
      Scenario & User Type & \#Tasks & \#Users & \#Conv. \\
      \midrule
        \multirow{2}{*}{All} & Agent & 22 & 85 & 1856 \\
        & Human & 79 & 146 & 2124 \\
      \midrule
        \multirow{2}{*}{\shortstack{Travel \\ Planning}} & Agent & 8 & 85 & 557 \\
        & Human & 31 & 146 & 579 \\
      \midrule
        \multirow{2}{*}{\shortstack{Recipe \\ Planning}} & Agent & 4 & 85 & 352 \\
        & Human & 12 & 138 & 470 \\
      \midrule
        \multirow{2}{*}{\shortstack{Preparing \\ Gifts}} & Agent & 5 & 85 & 474 \\
        & Human & 18 & 140 & 603 \\
      \midrule
        \multirow{2}{*}{\shortstack{Skills Learning \\ Planning}} & Agent & 5 & 85 & 473 \\
        & Human & 18 & 140 & 472 \\
      \bottomrule
    \end{tabular}
  \label{tab:dataset_basic}
\end{table}

%% file: Tables/manual.tex
\begin{table}[!ht]
    \centering
    \caption{Manual verification results of GPT-4o’s automatic classification.}
    \begin{tabular}{cc|c}
        \toprule
        \textbf{Dimensions} & \textbf{Sub-dim.} & \textbf{Match Rate} \\
        \midrule
        \textbf{All-in-one / Step-by-step} & / & $0.960$ \\
        \midrule
        \textbf{Question Broadness} & / & $0.890$ \\
        \midrule
        \textbf{Breadth-First / Depth-First} & / & $0.920$ \\
        \midrule
        \textbf{Context Dependency} & / & $0.900$ \\
        \midrule
        \multirow{2}{*}{\textbf{Feedback}} & \textbf{Feedback} & $0.980$ \\
        & \textbf{Promise} & $0.970$ \\
        \midrule
        \multirow{2}{*}{\textbf{Language Style}} & \textbf{Politeness} & $0.995$ \\
        & \textbf{Formality} & $0.930$ \\
        \midrule
        \multirow{2}{*}{\textbf{Usefulness}} & \textbf{Utility} & $0.965$ \\
        & \textbf{Operability} & $0.940$ \\
        \bottomrule
    \end{tabular}
    \label{tab:manual_eval}
\end{table}

%% file: Sections/4.tex
\section{Findings: Conversation Strategy}
\label{Section-4}

\subsection{All-in-one or Step-by-step}


We categorize users' problem-solving approaches during LLM interactions into two distinct 
types:
\begin{itemize}[leftmargin=*, parsep=2pt]
    \item \textbf{All-in-One}: Users adopting this strategy explicitly request a comprehensive plan at the outset, seeking a structured solution framework before addressing finer details.
    This top-down approach reflects a preference for upfront clarity, often observed in LLM simulators that demand complete task specifications early in the conversation.
    \item \textbf{Step-by-Step}: Users following this strategy iteratively refine their solutions through incremental exchanges, dynamically adjusting their requests without articulating an overarching plan initially.
    This bottom-up approach is more prevalent among human users, where tasks are decomposed progressively.
\end{itemize}

\begin{figure}[h]
    \centering
    \includegraphics[width=0.9\linewidth]{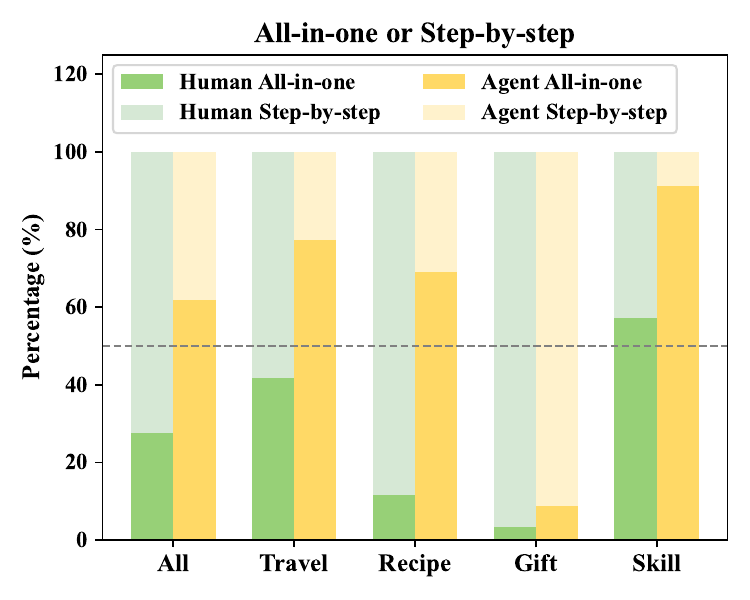}
    \caption{The distribution of all-in-one or step-by-step dimension on our constructed conversations. The dashed line indicates the 50\% threshold.}
    \Description[<short description>]{<long description>}
    \label{fig:all-step}
\end{figure}

As evidenced by Figure~\ref{fig:all-step}, human users favor the step-by-step significantly more than agent users, who employ more all-in-one queries except in the preparing gifts scenario.
This contrast underscores fundamental differences in planning styles: agents prioritize holistic solutions, while humans often blend planning with iterative refinement.

\textbf{\textit{Finding:} Human users prefer step-by-step strategy than agent users.}

\subsection{Question Broadness}

\input{Tables/question}

\textbf{Question broadness} refers to the granularity of user questions.
We employed a five-level scale, where 3 represents neutral, 1-2 corresponds to broad queries, and 4-5 indicates specific questions.
The distribution of automated annotation scores is presented in Table~\ref{tab:question} (a).

Generally, human users tend to ask more specific questions in most scenarios.
However, in the skills learning planning scenario, LLM agent users surprisingly asked more specific questions.
Task requirements, such as learning programming languages, require participants to envision a situation where they need to acquire a new skill.
Human users, being unfamiliar with the skills they're supposed to learn, may find it difficult to pose very specific questions.
This difference highlights the significance of LLM simulation in such scenarios.

\textbf{\textit{Finding:} Human users tend to ask questions with higher specificity in most scenarios.}

\subsection{Breadth-First or Depth-First}

\input{Tables/BF-DF}

When formulating plans in various scenarios, users can continuously optimize and improve based on different aspects of the plan (Breadth) and conduct multiple rounds of specific refinements for individual aspects (Depth).
Therefore, we have designed the \textit{Breadth-First or Depth-First} classification based on the depth and breadth demonstrated by users in the conversation:
\begin{itemize}[leftmargin=*, parsep=2pt]
    \item \textbf{Depth-First} (DF): The user focuses intensely on a specific aspect of their plan, thoroughly exploring it before moving on to another aspect.
    \item \textbf{Breadth-First} (BF): The user considers a wide range of aspects of their plan, exploring each one briefly without going into much detail.
    \item \textbf{Depth-First, Then Breadth} (DF-B): The user initially examines a particular aspect of their plan in great detail before expanding their inquiry to include a broader range of aspects.
    \item \textbf{Breadth-First, Then Depth} (BF-D): The user begins by covering a wide array of aspects superficially and subsequently chooses specific aspects to explore in detail.
\end{itemize}

Agent users showed a similar distribution to human users on this dimension, except in the preparing gifts scenario.
Almost every conversation begins by first establishing a comprehensive plan before considering further refinement and optimization for specific aspects (\textit{BF-D} and \textit{BF}).
However, in the preparing gifts scenario, agent users exhibit a higher proportion of instances where they omit refinement of the selected gift option (\textit{BF}).
This disparity constitutes the primary difference between agent users and human users in this dimension.

\textbf{\textit{Finding:} Agent users showed similar distributions with human users on the breadth-first or depth-first dimension.}

\section{Findings: Interaction Style}
\label{Section-5}

\subsection{User Engagement}

User engagement in conversational interactions can be characterized by two primary metrics: the number of conversational turns and the average tokens per turn.
These measures offer insights into how users engage in conversations and articulate their needs.
\input{Tables/conv_length}

Our analysis reveals distinct patterns between human users and LLM agents.
As shown in Table~\ref{tab:conv_length}, LLM agents typically engage in fewer conversational turns but generate significantly longer responses per turn compared to human users across all scenarios.
This pattern suggests that simulated interactions tend to prioritize comprehensive, single-turn responses, while human conversations more commonly involve iterative exchanges with shorter turns.

Particularly in the \textit{Preparing Gifts} and \textit{Skills Learning Planning} scenarios, we observe that agent interactions involve substantially fewer turns than human conversations.
This difference indicates that current simulations may not fully capture the comprehensive, multi-turn nature of human problem-solving approaches.
Furthermore, human users demonstrate more efficient communication, achieving higher assistant token counts with relatively shorter inputs, which suggests their ability to express detailed needs more concisely.

The observed patterns highlight fundamental differences in engagement styles: human users tend toward more interactive conversations with balanced turn lengths.
At the same time, LLM agents favor condensed interactions with more verbose individual turns.

\textbf{\textit{Finding:} Agent users tend to give more information in a single turn, while humans prefer to interact in more turns.}

\subsection{Context Dependency}

\input{Tables/context}

\textbf{Context dependency} refers to how much the user's questions depend on prior context or information in the conversation.
We similarly adopted a five-level scale, where 1-2 suggests low relevance, and 4-5 represents high relevance.

The annotation results in Table~\ref{tab:context} (b) reveal behavioral differences between agent users and human users.
In general, human users tend to ask questions with higher contextual relevance.
This is primarily reflected in the higher proportion of level-4 scores among human users, while agents predominantly receive level-3 scores.

\textbf{\textit{Finding:} Human users generally ask more relevant questions with contextual information than agent users.}

\subsection{Feedback Polarity and Promise}

\input{Tables/feedback}

\textbf{Feedback polarity} refers to how the user responds to the assistant's responses.
We analyzed user feedback on assistant responses in each conversation, categorizing them into four types: \textit{positive}, \textit{negative}, \textit{no} feedback, and \textit{both} positive and negative.
Table~\ref{tab:feedback} presents the distribution of user feedback polarity.
LLM agent users predominantly provided positive feedback in almost all cases, while human users more frequently chose not to give explicit feedback.
Additionally, human users exhibited a certain proportion of the other three types of feedback responses.

\textbf{Promise} is determined by whether users explicitly affirm that they will follow through on the assistant's suggestions.
This signal reveals how users position the assistant's role, i.e., whether as an authoritative guide to be acted upon or simply an advisory presence.
As shown in Table~\ref{tab:feedback}, agent users frequently promise to adopt the assistant's recommendations, whereas human users rarely indicate such intentions.
This further indicates that agent users treat assistants as command issuers and themselves as service providers.
This difference is especially prominent in the \textit{Skills Learning Planning} scenario, where agent users often close with phrases like "I will do that" or "Thanks, I'll follow your plan."
In contrast, human users typically move on after receiving advice without expressing any commitment to action.

\textbf{\textit{Finding:} Agent users almost always give positive feedback, and frequently promise to adopt the assistant's recommendations, while human users do not show these behaviors.}

\subsection{Language Style}

\input{Tables/language_style}

LLM agent users and real human users tend to show different language styles in their conversations.
\textbf{Politeness} is determined by the user's choice of language tone when interacting with the assistant.
We divided this attribute into three categories: Polite, Neutral, and Impolite.
This aspect captures how considerate or socially softened users are in expressing their requests and engaging in conversations.
As shown in Table~\ref{tab:language_style}, agent users consistently display polite behavior across all scenarios, with no instances of neutral or impolite tone.
Human users, on the other hand, are far more likely to use a neutral tone, with polite expressions appearing far less frequently and impolite ones only in rare cases. 
This suggests that LLM-simulated users are designed to follow overly courteous conventions, further proving that they mirror assistant-like behavior rather than natural user interaction.
Human users, by contrast, may see the assistant as a tool and prioritize clarity and efficiency over politeness.

\textbf{Formality} is assessed based on the user's linguistic register, i.e., whether their language resembles formal written expression or informal, oral communication.
We divide this attribute into three categories: Oral, Formal, and Mixed.
This aspect helps illuminate the stylistic tendencies users adopt when engaging with an assistant in task-oriented conversations.
As shown in Table~\ref{tab:language_style}, agent users predominantly use a formal tone across all scenarios, while human users exhibit oral and formal language styles in roughly equal proportions during conversations.
In specific contexts, users tend to adopt a more formal style in skill learning planning scenarios, whereas they favor an oral style in travel planning scenarios.
This suggests that the characteristics of the scenarios significantly influence the language styles of human users.

\input{Tables/satisfaction}

\textbf{\textit{Finding:} Agent users are always polite during conversations, while human users will show different language styles based on the suggestions they get.}

\section{Findings: Conversation Evaluation}
\label{Section-6}

\subsection{User Satisfaction}
\label{Section-satisfaction}

We evaluated user satisfaction across three sub-dimensions (Detail, Practical Utility, and Diversity), each scored on a 0-2 scale:
\begin{itemize}[leftmargin=*, parsep=2pt]
    \item \textbf{Detail} assessed response comprehensiveness (0 for incomplete, 1 for partially detailed, 2 for fully actionable).
    \item \textbf{Practical Utility} measured implementation feasibility (0 for unrealistic, 1 for partially actionable, 2 for readily executable).
    \item \textbf{Diversity} quantified solution variety (0 for single option, 1 for limited alternatives, 2 for multiple approaches).
\end{itemize}

For human users, satisfaction scores were collected via post-conversation questionnaires.
Agent users were evaluated through dual methods: LLM self-reports and third-party annotations.
As shown in Table~\ref{tab:satisfaction}, LLM self-reports exhibited strong positivity bias (mean scores over 1.9 across all sub-dimensions), suggesting limited reliability without proper controls.

Agent users showed significantly lower \textit{Diversity} scores (particularly in \textit{Recipe Planning}), indicating greater acceptance of single solutions compared to humans.
Human users prioritized \textit{Practical Utility} (scoring lower than other dimensions except in Recipe Planning), demonstrating stricter real-world feasibility requirements.
The discrepancy highlights fundamental behavioral differences: humans demand executable solutions, while agents exhibit weaker diversity expectations in certain scenarios.

\textbf{\textit{Finding:} Human users prioritized practical utility during conversations.}

\subsection{Hallucination Awareness}

To evaluate response reliability, we systematically measured hallucination frequency in assistant responses through post-conversation annotation.
For human-user conversations, participants directly flagged hallucinatory content in each assistant's turn.
Third-party annotators performed equivalent evaluations for agent-user conversations due to the same reason mentioned in Section~\ref{Section-satisfaction}.

\begin{figure}[h]
    \centering
    \includegraphics[width=\linewidth]{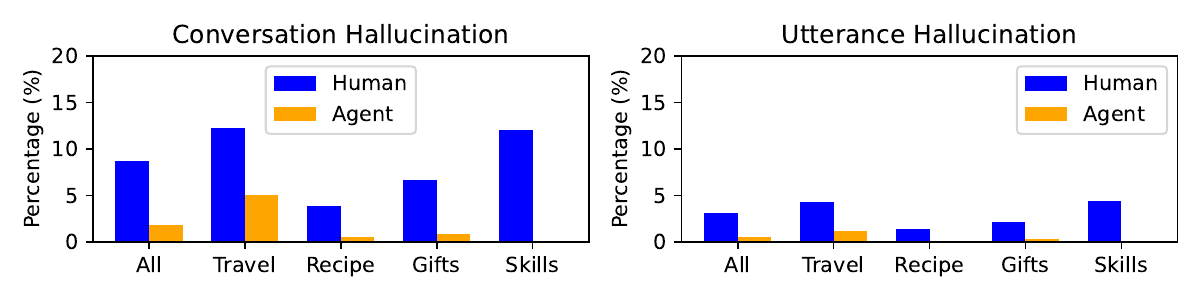}
    \caption{Distribution of hallucination detected on our constructed conversations. We report the rate both on the conversation level and the utterance level.}
    \Description[<short description>]{<long description>}
    \label{fig:hallucination}
    \vspace{-1em}
\end{figure}

As illustrated in Figure~\ref{fig:hallucination}, we present hallucination rates at both conversation-level (per conversation) and turn-level (per utterance) granularities.
We found significantly higher hallucination prevalence in human conversations compared to agent-user interactions across both metrics.
Human users' tendency for granular inquiries (e.g., requesting specific weather forecasts or skill-learning URLs) frequently exposed knowledge boundaries, triggering more hallucinatory responses when assistant capabilities were exceeded.

\textbf{\textit{Finding:} Human users are more likely to detect hallucinations in assistant responses during a conversation.}

\subsection{Usefulness}

\input{Tables/usefulness}

The Usefulness dimension evaluates the practical value of an LLM's suggestions during user interactions, consisting of two sub-dimensions:
\begin{itemize}[leftmargin=*, parsep=2pt]
    \item \textbf{Utility} measures alignment with user intent and task-solving effectiveness. High-utility responses directly address needs (e.g., precise recommendations), while low-utility ones are off-topic or generic.
    \item \textbf{Operability} assesses actionability, with high-operability responses offering concrete steps (e.g., "Book a 7 PM sushi class") and low-operability ones being vague (e.g., "Explore local culture").
\end{itemize}
The above two sub-dimensions are divided into three levels through automated analysis: high, moderate, and low.

As shown in Table~\ref{tab:usefulness}, both \textbf{Utility} and \textbf{Operability} scores are overwhelmingly "High" across all scenarios.
Human users exhibit slightly higher \textbf{Operability} in \textit{Travel Planning}, while agent users excel in \textit{Preparing Gifts}.
In general, human users place great importance on the usefulness of the overall plan during the conversation.
Agent users who have been guided through task requirement prompting can also achieve similar performance.

\textbf{\textit{Finding:} Agent users show a similar distribution with human users on the usefulness dimension.}

%% file: Tables/question.tex
\begin{table}[htbp]
    \centering
    \caption{Distribution of the scores of question broadness on our constructed conversations.}
    \resizebox{0.9\linewidth}{!}{
    \begin{tabular}{cc|cccccc}
        \toprule
        \multirow{2}{*}{Scenario} & \multirow{2}{*}{Source} & \multicolumn{6}{|c}{\textbf{Question Broadness}} \\ 
         &  & 1 & 2 & 3 & 4 & 5 & Avg. \\ 
        \midrule
        \multirow{2}{*}{All} & Agent & 0.00\% & 3.39\% & 43.05\% & 53.50\% & 0.05\% & 3.50 \\ 
         & Human & 0.00\% & 2.21\% & 28.01\% & 67.33\% & 2.45\% & 3.70 \\ 
        \midrule
        \multirow{2}{*}{\shortstack{Travel \\ Planning}} & Agent & 0.00\% & 10.23\% & 59.78\% & 29.80\% & 0.18\% & 3.20 \\ 
         & Human & 0.00\% & 3.97\% & 34.54\% & 57.17\% & 4.32\% & 3.62 \\ 
        \midrule
        \multirow{2}{*}{\shortstack{Recipe \\ Planning}} & Agent & 0.00\% & 1.70\% & 55.11\% & 43.18\% & 0.00\% & 3.41 \\ 
         & Human & 0.00\% & 1.28\% & 28.94\% & 67.02\% & 2.77\% & 3.71 \\ 
        \midrule
        \multirow{2}{*}{\shortstack{Preparing \\ Gifts}} & Agent & 0.00\% & 0.00\% & 40.72\% & 59.28\% & 0.00\% & 3.59 \\ 
         & Human & 0.00\% & 0.50\% & 16.25\% & 81.92\% & 1.33\% & 3.84 \\ 
        \midrule
        \multirow{2}{*}{\shortstack{Skills Learning \\ Planning}} & Agent & 0.00\% & 0.00\% & 16.70\% & 83.30\% & 0.00\% & 3.83 \\ 
         & Human & 0.00\% & 3.18\% & 34.11\% & 61.44\% & 1.27\% & 3.61 \\ 
        \bottomrule
    \end{tabular}}
    \label{tab:question}
\end{table}

%% file: Tables/BF-DF.tex
\begin{table}[htbp]
  \centering
  \caption{Distribution of depth-first vs. breadth-first dimension on two data sources.}
  \begin{threeparttable}[c]
  \resizebox{0.9\linewidth}{!}{
    \begin{tabular}{cc|cccc}
      \toprule
      \multirow{2}{*}{Scenario} & \multirow{2}{*}{Source} & \multicolumn{4}{|c}{Depth Vs. Breadth} \\
      & & DF & BF & DF-B & BF-D \\
      \midrule
      \multirow{2}{*}{All} & Agent & 0.27\% & 15.41\% & 0.22\% & 84.11\% \\
      & Human & 1.41\% & 3.39\% & 0.71\% & 94.49\% \\
      \midrule
      \multirow{2}{*}{\shortstack{Travel \\ Planning}} & Agent & 0.00\% & 0.72\% & 0.18\% & 99.10\% \\
      & Human & 0.35\% & 3.28\% & 0.69\% & 95.68\% \\
      \midrule
      \multirow{2}{*}{\shortstack{Recipe \\ Planning}} & Agent & 0.28\% & 2.27\% & 0.00\% & 97.44\% \\
      & Human & 1.49\% & 2.13\% & 0.43\% & 95.96\% \\
      \midrule
      \multirow{2}{*}{\shortstack{Preparing \\ Gifts}} & Agent & 0.84\% & 55.27\% & 0.63\% & 43.25\% \\
      & Human & 1.99\% & 5.64\% & 0.50\% & 91.87\% \\
      \midrule
      \multirow{2}{*}{\shortstack{Skills Learning \\ Planning}} & Agent & 0.00\% & 2.54\% & 0.00\% & 97.46\% \\
      & Human & 1.91\% & 1.91\% & 1.27\% & 94.92\% \\
      \bottomrule
    \end{tabular}
  }
  \end{threeparttable}
  \label{tab:bf-df}
\end{table}

%% file: Tables/conv_length.tex
\begin{table}[htbp]
  \centering
  \caption{The length of conversations under different scenarios from LLM agent users and human users. "T/T" represents the average number of tokens per turn in the conversation, where a turn refers to one complete user query and assistant response cycle.}
  \resizebox{0.9\linewidth}{!}{
    \begin{tabular}{cc|ccccccc}
      \toprule
      \multirow{3}{*}{Scenario} & \multirow{3}{*}{Source} & \multirow{3}{*}{\shortstack{Avg. \\ Turns}} & \multirow{3}{*}{\shortstack{Avg. \\ User \\ Tokens}} & \multirow{3}{*}{\shortstack{Avg. \\ User \\ T/T}} & \multirow{3}{*}{\shortstack{Avg. \\ Assistant \\ T/T}} \\
      & & & & \\
      & & & & \\
      \midrule
        \multirow{2}{*}{All} & Agent & 3.82 & 679.20 & 177.65 & 496.21 \\
        & Human & 4.44 & 168.44 & 37.47 & 1307.93 \\
      \midrule
        \multirow{2}{*}{\shortstack{Travel \\ Planning}} & Agent & 5.00 & 1056.01 & 211.20 & 567.52 \\
        & Human & 5.07 & 209.28 & 40.80 & 1504.76 \\
      \midrule
        \multirow{2}{*}{\shortstack{Recipe \\ Planning}} & Agent & 4.14 & 694.70 & 167.72 & 448.66 \\
        & Human & 4.21 & 156.51 & 36.82 & 1271.28 \\
      \midrule
        \multirow{2}{*}{\shortstack{Preparing \\ Gifts}} & Agent & 3.22 & 395.86 & 123.04 & 347.84 \\
        & Human & 4.28 & 144.67 & 33.36 & 972.16 \\
      \midrule
        \multirow{2}{*}{\shortstack{Skills Learning \\ Planning}} & Agent & 2.81 & 507.89 & 180.90 & 568.86 \\
        & Human & 4.11 & 160.57 & 38.59 & 1493.54 \\
      \bottomrule
    \end{tabular}
  }
  \label{tab:conv_length}
\end{table}

%% file: Tables/context.tex
\begin{table}[h]
    \centering
    \caption{Distribution of the scores of context dependency on our constructed conversations.}
    \resizebox{0.9\linewidth}{!}{
    \begin{tabular}{cc|cccccc}
        \toprule
        \multirow{2}{*}{Scenario} & \multirow{2}{*}{Source} & \multicolumn{6}{|c}{\textbf{Context Dependency}} \\ 
         &  & 1 & 2 & 3 & 4 & 5 & Avg. \\ 
        \midrule
        \multirow{2}{*}{All} & Agent & 0.32\% & 3.66\% & 79.63\% & 16.27\% & 0.11\% & 3.12  \\ 
         & Human & 0.33\% & 3.48\% & 51.27\% & 41.71\% & 3.20\% & 3.44  \\ 
        \midrule
        \multirow{2}{*}{\shortstack{Travel \\ Planning}} & Agent & 0.00\% & 0.00\% & 63.20\% & 36.45\% & 0.36\% & 3.37  \\ 
         & Human & 0.00\% & 2.59\% & 42.31\% & 48.19\% & 6.91\% & 3.59  \\ 
        \midrule
        \multirow{2}{*}{\shortstack{Recipe \\ Planning}} & Agent & 0.00\% & 0.85\% & 85.80\% & 13.35\% & 0.00\% & 3.13  \\ 
         & Human & 0.64\% & 2.55\% & 52.77\% & 41.70\% & 2.34\% & 3.43  \\ 
        \midrule
        \multirow{2}{*}{\shortstack{Preparing \\ Gifts}} & Agent & 1.27\% & 12.87\% & 81.22\% & 4.64\% & 0.00\% & 2.89  \\ 
         & Human & 0.66\% & 2.49\% & 50.75\% & 44.61\% & 1.49\% & 3.44  \\ 
        \midrule
        \multirow{2}{*}{\shortstack{Skills Learning \\ Planning}} & Agent & 0.00\% & 0.85\% & 92.81\% & 6.34\% & 0.00\% & 3.05  \\ 
         & Human & 0.00\% & 6.78\% & 61.44\% & 30.108\% & 1.69\% & 3.27  \\ 
        \bottomrule
    \end{tabular}}
    \label{tab:context}
\end{table}

%% file: Tables/feedback.tex

\begin{table}[ht]
  \centering
  \caption{Distribution of different feedback polarity and promise on our constructed conversations.}
  \begin{threeparttable}[c]
  \resizebox{0.9\linewidth}{!}{
    \begin{tabular}{cc|cccc|cc}
      \toprule
      \multirow{2}{*}{Scenario} & \multirow{2}{*}{Source} & \multicolumn{4}{|c}{Feedback} & \multicolumn{2}{|c}{Promise} \\
      & & No & Pos. & Neg. & Both & Yes & No \\
      \midrule
      \multirow{2}{*}{All} & Agent & 0.32\% & 99.62\% & 0.00\% & 0.05\% & 95.58\% & 4.42\% \\
      & Human & 71.66\% & 25.05\% & 1.98\% & 1.32\% & 13.61\% & 86.39\% \\
      \midrule
      \multirow{2}{*}{\shortstack{Travel \\ Planning}} & Agent & 0.18\% & 99.82\% & 0.00\% & 0.00\% & 94.79\% & 5.21\% \\
      & Human & 64.68\% & 30.43\% & 2.55\% & 2.34\% & 15.74\% & 84.26\% \\
      \midrule
      \multirow{2}{*}{\shortstack{Recipe \\ Planning}} & Agent & 0.57\% & 99.15\% & 0.00\% & 0.28\% & 88.92\% & 11.08\% \\
      & Human & 74.09\% & 23.49\% & 1.21\% & 1.21\% & 14.16\% & 85.84\% \\
      \midrule
      \multirow{2}{*}{\shortstack{Preparing \\ Gifts}} & Agent & 0.63\% & 99.37\% & 0.00\% & 0.00\% & 97.68\% & 2.32\% \\
      & Human & 68.33\% & 28.69\% & 2.16\% & 0.83\% & 17.91\% & 82.09\% \\
      \midrule
      \multirow{2}{*}{\shortstack{Skills Learning \\ Planning}} & Agent & 0.00\% & 100.00\% & 0.00\% & 0.00\% & 99.37\% & 0.63\% \\
      & Human & 79.87\% & 16.95\% & 2.12\% & 1.06\% & 5.30\% & 94.70\% \\
      \bottomrule
    \end{tabular}
  }
  \end{threeparttable}
  \label{tab:feedback}
\end{table}

%% file: Tables/language_style.tex
\begin{table}[ht]
  \centering
  \caption{Distribution of user language styles on our constructed conversations.}
  \begin{threeparttable}[c]
  \resizebox{0.9\linewidth}{!}{
    \begin{tabular}{cc|ccc|ccc}
      \toprule
      \multirow{2}{*}{Scenario} & \multirow{2}{*}{Source} & \multicolumn{3}{|c}{Politeness} & \multicolumn{3}{|c}{Formality} \\
      & & Polite & Neutral & Impolite & Oral & Formal & Mixed \\
      \midrule
      \multirow{2}{*}{All} & Agent & 99.84\% & 0.16\% & 0.00\% & 12.98\% & 77.86\% & 9.16\% \\
      & Human & 4.71\% & 95.15\% & 0.14\% & 49.11\% & 49.81\% & 1.08\% \\
      \midrule
      \multirow{2}{*}{\shortstack{Travel \\ Planning}} & Agent & 99.64\% & 0.36\% & 0.00\% & 11.85\% & 72.89\% & 15.26\% \\
      & Human & 6.60\% & 93.40\% & 0.00\% & 72.13\% & 26.60\% & 1.28\% \\
      \midrule
      \multirow{2}{*}{\shortstack{Recipe \\ Planning}} & Agent & 100.00\% & 0.00\% & 0.00\% & 11.93\% & 83.52\% & 4.55\% \\
      & Human & 3.97\% & 95.68\% & 0.35\% & 42.31\% & 56.82\% & 0.86\% \\
      \midrule
      \multirow{2}{*}{\shortstack{Preparing \\ Gifts}} & Agent & 99.79\% & 0.21\% & 0.00\% & 23.63\% & 69.62\% & 6.75\% \\
      & Human & 4.48\% & 95.52\% & 0.00\% & 54.56\% & 44.61\% & 0.83\% \\
      \midrule
      \multirow{2}{*}{\shortstack{Skills Learning \\ Planning}} & Agent & 100.00\% & 0.00\% & 0.00\% & 4.44\% & 87.74\% & 7.82\% \\
      & Human & 4.03\% & 95.76\% & 0.21\% & 27.54\% & 70.97\% & 1.48\% \\
      \bottomrule
    \end{tabular}
  }
  \end{threeparttable}
  \label{tab:language_style}
\end{table}

%% file: Tables/satisfaction.tex
\begin{table*}[ht]
  \centering
  \caption{Average scores and score distribution for detail level, practical utility, and diversity. For all metrics, higher scores indicate better quality. "TP" and "SR" represent third-party qualified annotators and self-reports from agent users, respectively.}
  \resizebox{0.8\linewidth}{!}{
  \begin{tabular}{cc|cccc|cccc|cccc}
    \toprule
    \multirow{2}{*}{Scenario} & \multirow{2}{*}{Source} & \multicolumn{4}{|c}{Detail Level}	& \multicolumn{4}{|c}{Practical Utility}  & \multicolumn{4}{|c}{Diversity}	\\
    & & 0 &	1 &	2 &	Avg. &	0 &	1 &	2 &	Avg. &	0 &	1 &	2 &	Avg.\\
    \midrule
    \multirow{3}{*}{All}
    & Agent-TP & 1.56\% & 9.54\% & 88.90\% & 1.873 & 0.81\% & 16.43\% & 82.76\% & 1.820 & 12.66\% & 13.31\% & 74.03\% & 1.614 \\
    & Agent-SR & 0.00\% & 0.32\% & 99.68\% & 1.997 & 0.00\% & 0.32\% & 99.68\% & 1.997 & 0.05\% & 9.05\% & 90.89\% & 1.908 \\
    & Human & 3.20\% & 28.53\% & 68.27\% & 1.651 & 1.79\% & 39.08\% & 59.13\% & 1.573 & 3.01\% & 27.82\% & 69.16\% & 1.661 \\
    \midrule
    \multirow{3}{*}{\shortstack{Travel \\ Planning}}
    & Agent-TP & 3.23\% & 14.00\% & 82.76\% & 1.795 & 2.15\% & 22.08\% & 75.76\% & 1.736 & 5.92\% & 17.41\% & 76.66\% & 1.707 \\
    & Agent-SR & 0.00\% & 0.36\% & 99.64\% & 1.996 & 0.00\% & 0.36\% & 99.64\% & 1.996 & 0.00\% & 2.69\% & 97.31\% & 1.973 \\
    & Human & 3.45\% & 30.40\% & 66.15\% & 1.627 & 1.38\% & 43.18\% & 55.44\% & 1.541 & 3.45\% & 28.15\% & 68.39\% & 1.649 \\
    \midrule
    \multirow{3}{*}{\shortstack{Recipe \\ Planning}}
    & Agent-TP & 3.12\% & 10.80\% & 86.08\% & 1.830 & 0.85\% & 18.75\% & 80.40\% & 1.795 & 52.27\% & 30.11\% & 17.61\% & 0.653 \\
    & Agent-SR & 0.00\% & 1.14\% & 98.86\% & 1.989 & 0.00\% & 1.14\% & 98.86\% & 1.989 & 0.28\% & 40.62\% & 59.09\% & 1.588 \\
    & Human & 2.98\% & 22.55\% & 74.47\% & 1.715 & 1.06\% & 26.60\% & 72.34\% & 1.713 & 2.55\% & 25.53\% & 71.91\% & 1.694 \\
    \midrule
    \multirow{3}{*}{\shortstack{Preparing \\ Gifts}}
    & Agent-TP & 0.00\% & 8.02\% & 91.98\% & 1.920 & 0.00\% & 13.29\% & 86.71\% & 1.867 & 0.00\% & 1.05\% & 98.95\% & 1.989 \\
    & Agent-SR & 0.00\% & 0.00\% & 100.00\% & 2.000 & 0.00\% & 0.00\% & 100.00\% & 2.000 & 0.00\% & 0.00\% & 100.00\% & 2.000 \\
    & Human & 3.32\% & 31.18\% & 65.51\% & 1.622 & 2.32\% & 40.13\% & 57.55\% & 1.552 & 3.15\% & 28.03\% & 68.82\% & 1.657 \\
    \midrule
    \multirow{3}{*}{\shortstack{Skills \\ Learning \\ Planning}}
    & Agent-TP & 0.00\% & 4.86\% & 95.14\% & 1.951 & 0.00\% & 11.21\% & 88.79\% & 1.888 & 3.81\% & 8.25\% & 87.95\% & 1.841 \\
    & Agent-SR & 0.00\% & 0.00\% & 100.00\% & 2.000 & 0.00\% & 0.00\% & 100.00\% & 2.000 & 0.00\% & 2.11\% & 97.89\% & 1.979 \\
    & Human & 2.97\% & 28.81\% & 68.22\% & 1.653 & 2.33\% & 45.13\% & 52.54\% & 1.502 & 2.75\% & 29.45\% & 67.80\% & 1.650 \\
    \bottomrule
  \end{tabular}
  }
  \label{tab:satisfaction}
\end{table*}

%% file: Tables/usefulness.tex
\begin{table}[ht]
  \centering
  \caption{Distribution of utility and operability on our constructed conversations.}
  \begin{threeparttable}[c]
  \resizebox{0.9\linewidth}{!}{
    \begin{tabular}{cc|ccc|ccc}
      \toprule
      \multirow{2}{*}{Scenario} & \multirow{2}{*}{Source} & \multicolumn{3}{|c}{Utility} & \multicolumn{3}{|c}{Operability} \\
      & & High & Moderate & Low & High & Moderate & Low \\
      \midrule
      \multirow{2}{*}{All} & Agent & 99.46\% & 0.48\% & 0.05\% & 95.47\% & 4.47\% & 0.05\% \\
      & Human & 98.54\% & 1.37\% & 0.09\% & 96.28\% & 3.63\% & 0.09\% \\
      \midrule
      \multirow{2}{*}{\shortstack{Travel \\ Planning}} & Agent & 99.10\% & 0.90\% & 0.00\% & 89.23\% & 10.77\% & 0.00\% \\
      & Human & 96.72\% & 3.28\% & 0.00\% & 94.30\% & 5.70\% & 0.00\% \\
      \midrule
      \multirow{2}{*}{\shortstack{Recipe \\ Planning}} & Agent & 98.86\% & 0.85\% & 0.28\% & 98.30\% & 1.42\% & 0.28\% \\
      & Human & 99.36\% & 0.43\% & 0.21\% & 98.94\% & 0.85\% & 0.21\% \\
      \midrule
      \multirow{2}{*}{\shortstack{Preparing \\ Gifts}} & Agent & 100.00\% & 0.00\% & 0.00\% & 96.41\% & 3.59\% & 0.00\% \\
      & Human & 99.34\% & 0.50\% & 0.17\% & 95.02\% & 4.81\% & 0.17\% \\
      \midrule
      \multirow{2}{*}{\shortstack{Skills Learning \\ Planning}} & Agent & 99.79\% & 0.21\% & 0.00\% & 99.79\% & 0.21\% & 0.00\% \\
      & Human & 98.94\% & 1.06\% & 0.00\% & 97.67\% & 2.33\% & 0.00\% \\
      \bottomrule
    \end{tabular}
  }
  \end{threeparttable}
  \label{tab:usefulness}
\end{table}

%% file: Sections/5.tex
\section{Summary of Findings and Discussion}

\subsection{Observation}

Overall, our comparative analysis of human and agent users reveals the following behavior differences:
\begin{enumerate}[leftmargin=*, parsep=2pt]
    \item \textbf{LLM agents} prefer a \textbf{All-in-one} strategy when developing plans except for \textbf{Preparing Gifts}, while \textbf{human users} are more likely to adopt an \textbf{Step-by-step} strategy except for \textbf{Skills Learning Planning}.
    \item In task-oriented conversations, \textbf{human users} tend to ask questions with \textbf{higher contextual relevance and specificity}.
    However, in some scenarios (such as skills learning planning), \textbf{LLM agents} are more adept at posing \textbf{specific} questions.
    \item Compared to human interactions, \textbf{LLM agents} produce \textbf{fewer but more verbose} conversation turns.
    \item \textbf{LLM agents} almost always \textbf{give positive feedback} and are almost always \textbf{polite} in conversations, whereas \textbf{human users rarely give any feedback}.
    \item \textbf{Agent users} tend to give \textbf{promise} for assistant's suggestions, while humans are significantly less likely to do so.
    \item \textbf{Human} prioritized \textbf{Practical Utility} and \textbf{Operability}, demonstrating stricter real-world feasibility requirements.
\end{enumerate}

Besides, we found several consistencies:
\begin{enumerate}[leftmargin=*, parsep=2pt]
    \item Both users showed similar distributions on the \textbf{breadth-first or depth-first dimension}.
    \item Human users place great importance on the \textbf{usefulness} of the overall plan during the conversation, as well as agent users.
\end{enumerate}

\subsection{Discussion}

In our experiments, the assistant LLMs for human user conversations encompassed multiple model series, while the agent user simulation process exclusively utilized GPT-series models.
Consequently, we conducted a comparative statistical analysis of the dimensional distributions between GPT-based and non-GPT-based assistant models within the human-user conversations.
Across the sub-dimensions of all-in-one or step-by-step, breadth-first or depth-first, context dependency, promise, politeness, utility, and operability, the distributions showed no statistically significant differences (p-value > 0.05).
For other sub-dimensions, although distributions varied among different LLMs acting as assistants, these variations relative to the agent user results maintained consistent patterns of difference and similarity. For instance:
\begin{itemize}[leftmargin=*, parsep=2pt]
    \item In the feedback polarity dimension, the option of providing no feedback was the most frequent choice among human users for both GPT-based (76.50\%) and non-GPT-based assistants (68.63\%).
    \item In the question broadness dimension, the mean scores for conversations involving GPT-based and non-GPT-based assistants were 3.75 and 3.67, respectively, both significantly higher than the agent user average of 3.50.
\end{itemize}

Therefore, the comparative analysis presented in this paper is based on the complete dataset, which encompasses conversations from both user types (human and agent).

\subsection{Limitations}

While this work provides comprehensive, multidimensional comparisons, it is essential to acknowledge several limitations.
First, our agent users exclusively employed GPT-series models, which may potentially limit the generalizability of our findings to user simulations powered by other LLM architectures (e.g., Claude) that may exhibit different behavioral characteristics.
Second, due to resource constraints, portions of our analysis relied on automated evaluations rather than full human annotation.
While we validated reliability through sampling (Section~\ref{Section-settings}), complete human annotation would strengthen conclusiveness.
Last, the four selected scenarios, though diverse, represent a subset of possible task-oriented interactions.
Behaviors in highly specialized domains (e.g., medical or legal consultations) may differ substantially.

\subsection{Potential Usage}

The findings and methodological framework presented in this study offer valuable insights for advancing LLM-based user simulation and conversational systems.
Based on our comparative analysis, future user simulation can be improved by refining diversity awareness, scenario-specific adaptation, and hallucination mitigation strategies.
Furthermore, our multi-aspect, multi-dimensional taxonomy provides a generalizable framework for analyzing user behavior patterns across task-oriented scenarios, enabling more precise personalized user modeling and targeted enhancements of conversational systems — such as optimizing response strategies to improve user satisfaction.
These contributions collectively support the development of more human-like conversational agents and data-driven improvements in human-AI interaction.

\section{Conclusion}

This study bridges a critical gap in conversational AI research by establishing the first systematic comparison between human users and LLM agent users in task-oriented conversations.
Through a novel three-aspect, ten-dimensional analytical framework, we enable fine-grained evaluation of user behaviors across controlled scenarios.
Our analysis reveals distinct and consistent characteristics of LLM-agent-simulated users compared to humans, providing fundamental insights to advance LLM-based user simulation methodologies.
These contributions collectively offer both a structured approach for future comparative studies and actionable guidance for developing more human-like conversational systems.

%% file: main.bib
@inproceedings{zhu2025llm,
  title={A llm-based controllable, scalable, human-involved user simulator framework for conversational recommender systems},
  author={Zhu, Lixi and Huang, Xiaowen and Sang, Jitao},
  booktitle={Proceedings of the ACM on Web Conference 2025},
  pages={4653--4661},
  year={2025}
}

@inproceedings{sanchez2024automating,
  title={Automating the development of task-oriented llm-based chatbots},
  author={S{\'a}nchez Cuadrado, Jes{\'u}s and P{\'e}rez-Soler, Sara and Guerra, Esther and De Lara, Juan},
  booktitle={Proceedings of the 6th ACM Conference on Conversational User Interfaces},
  pages={1--10},
  year={2024}
}

@article{horowitz2025llm,
  title={LLM Agents Display Human Biases but Exhibit Distinct Learning Patterns},
  author={Horowitz, Idan and Plonsky, Ori},
  journal={arXiv preprint arXiv:2503.10248},
  year={2025}
}

@article{huijzer2023large,
  title={Large language models show human behavior},
  author={Huijzer, Rik and Hill, Yannick},
  year={2023},
  publisher={PsyArXiv}
}

@article{alsagheer2024comparing,
  title={Comparing rationality between large language models and humans: Insights and open questions},
  author={Alsagheer, Dana and Karanjai, Rabimba and Diallo, Nour and Shi, Weidong and Lu, Yang and Beydoun, Suha and Zhang, Qiaoning},
  journal={arXiv preprint arXiv:2403.09798},
  year={2024}
}

@inproceedings{fontana2025nicer,
  title={Nicer Than Humans: How Do Large Language Models Behave in the Prisoner's Dilemma?},
  author={Fontana, Nicol{\'o} and Pierri, Francesco and Aiello, Luca Maria},
  booktitle={Proceedings of the International AAAI Conference on Web and Social Media},
  volume={19},
  pages={522--535},
  year={2025}
}

@inproceedings{xie2024can,
  title={Can Large Language Model Agents Simulate Human Trust Behavior?},
  author={Xie, Chengxing and Chen, Canyu and Jia, Feiran and Ye, Ziyu and Lai, Shiyang and Shu, Kai and Gu, Jindong and Bibi, Adel and Hu, Ziniu and Jurgens, David and others},
  booktitle={The Thirty-eighth Annual Conference on Neural Information Processing Systems},
  year={2024}
}

@article{jiaqi2025comparative,
  title={A comparative study of large language models and human personality traits},
  author={Jiaqi, Wang and others},
  journal={arXiv preprint arXiv:2505.14845},
  year={2025}
}

@article{sekulic2024reliable,
  title={Reliable LLM-based user simulator for task-oriented dialogue systems},
  author={Sekuli{\'c}, Ivan and Terragni, Silvia and Guimar{\~a}es, Victor and Khau, Nghia and Guedes, Bruna and Filipavicius, Modestas and Manso, Andr{\'e} Ferreira and Mathis, Roland},
  journal={arXiv preprint arXiv:2402.13374},
  year={2024}
}

@inproceedings{zhang2020evaluating,
  title={Evaluating conversational recommender systems via user simulation},
  author={Zhang, Shuo and Balog, Krisztian},
  booktitle={Proceedings of the 26th acm sigkdd international conference on knowledge discovery \& data mining},
  pages={1512--1520},
  year={2020}
}

@article{wang2023rethinking,
  title={Rethinking the evaluation for conversational recommendation in the era of large language models},
  author={Wang, Xiaolei and Tang, Xinyu and Zhao, Wayne Xin and Wang, Jingyuan and Wen, Ji-Rong},
  journal={arXiv preprint arXiv:2305.13112},
  year={2023}
}

@inproceedings{zhu2024reliable,
  title={How Reliable is Your Simulator? Analysis on the Limitations of Current LLM-based User Simulators for Conversational Recommendation},
  author={Zhu, Lixi and Huang, Xiaowen and Sang, Jitao},
  booktitle={Companion Proceedings of the ACM on Web Conference 2024},
  pages={1726--1732},
  year={2024}
}

@article{wang2023recagent,
  title={Recagent: A novel simulation paradigm for recommender systems},
  author={Wang, Lei and Zhang, Jingsen and Chen, Xu and Lin, Yankai and Song, Ruihua and Zhao, Wayne Xin and Wen, Ji-Rong},
  journal={arXiv preprint arXiv:2306.02552},
  year={2023}
}

@inproceedings{zhang2024agentcf,
  title={Agentcf: Collaborative learning with autonomous language agents for recommender systems},
  author={Zhang, Junjie and Hou, Yupeng and Xie, Ruobing and Sun, Wenqi and McAuley, Julian and Zhao, Wayne Xin and Lin, Leyu and Wen, Ji-Rong},
  booktitle={Proceedings of the ACM on Web Conference 2024},
  pages={3679--3689},
  year={2024}
}

@inproceedings{park2023generative,
  title={Generative agents: Interactive simulacra of human behavior},
  author={Park, Joon Sung and O'Brien, Joseph and Cai, Carrie Jun and Morris, Meredith Ringel and Liang, Percy and Bernstein, Michael S},
  booktitle={Proceedings of the 36th Annual ACM Symposium on User Interface Software and Technology},
  pages={1--22},
  year={2023}
}

@article{chan2023chateval,
  title={Chateval: Towards better llm-based evaluators through multi-agent debate},
  author={Chan, Chi-Min and Chen, Weize and Su, Yusheng and Yu, Jianxuan and Xue, Wei and Zhang, Shanghang and Fu, Jie and Liu, Zhiyuan},
  journal={arXiv preprint arXiv:2308.07201},
  year={2023}
}

@inproceedings{afzali2023usersimcrs,
  title={UserSimCRS: a user simulation toolkit for evaluating conversational recommender systems},
  author={Afzali, Jafar and Drzewiecki, Aleksander Mark and Balog, Krisztian and Zhang, Shuo},
  booktitle={Proceedings of the Sixteenth ACM International Conference on Web Search and Data Mining},
  pages={1160--1163},
  year={2023}
}

@inproceedings{sun2021simulating,
  title={Simulating user satisfaction for the evaluation of task-oriented dialogue systems},
  author={Sun, Weiwei and Zhang, Shuo and Balog, Krisztian and Ren, Zhaochun and Ren, Pengjie and Chen, Zhumin and de Rijke, Maarten},
  booktitle={Proceedings of the 44th International ACM SIGIR Conference on Research and Development in Information Retrieval},
  pages={2499--2506},
  year={2021}
}

@inproceedings{sekulic2022evaluating,
  title={Evaluating mixed-initiative conversational search systems via user simulation},
  author={Sekuli{\'c}, Ivan and Aliannejadi, Mohammad and Crestani, Fabio},
  booktitle={Proceedings of the Fifteenth ACM International Conference on Web Search and Data Mining},
  pages={888--896},
  year={2022}
}

@article{askari2023generating,
  title={Generating synthetic documents for cross-encoder re-rankers: A comparative study of chatgpt and human experts},
  author={Askari, Arian and Aliannejadi, Mohammad and Kanoulas, Evangelos and Verberne, Suzan},
  journal={arXiv preprint arXiv:2305.02320},
  year={2023}
}

@inproceedings{huang2023chatgpt,
  title={Is chatgpt better than human annotators? potential and limitations of chatgpt in explaining implicit hate speech},
  author={Huang, Fan and Kwak, Haewoon and An, Jisun},
  booktitle={Companion proceedings of the ACM web conference 2023},
  pages={294--297},
  year={2023}
}

@article{whitehouse2023llm,
  title={LLM-powered data augmentation for enhanced cross-lingual performance},
  author={Whitehouse, Chenxi and Choudhury, Monojit and Aji, Alham Fikri},
  journal={arXiv preprint arXiv:2305.14288},
  year={2023}
}

@inproceedings{hu2023unlocking,
  title={Unlocking the potential of user feedback: Leveraging large language model as user simulators to enhance dialogue system},
  author={Hu, Zhiyuan and Feng, Yue and Luu, Anh Tuan and Hooi, Bryan and Lipani, Aldo},
  booktitle={Proceedings of the 32nd ACM International Conference on Information and Knowledge Management},
  pages={3953--3957},
  year={2023}
}

@article{mok2024llm,
  title={LLM-based Frameworks for API Argument Filling in Task-Oriented Conversational Systems},
  author={Mok, Jisoo and Kachuee, Mohammad and Dai, Shuyang and Ray, Shayan and Taghavi, Tara and Yoon, Sungroh},
  journal={arXiv preprint arXiv:2407.12016},
  year={2024}
}

@article{zhang2019dialogpt,
  title={Dialogpt: Large-scale generative pre-training for conversational response generation},
  author={Zhang, Yizhe and Sun, Siqi and Galley, Michel and Chen, Yen-Chun and Brockett, Chris and Gao, Xiang and Gao, Jianfeng and Liu, Jingjing and Dolan, Bill},
  journal={arXiv preprint arXiv:1911.00536},
  year={2019}
}

@article{peng2021soloist,
  title={Soloist: Building task bots at scale with transfer learning and machine teaching},
  author={Peng, Baolin and Li, Chunyuan and Li, Jinchao and Shayandeh, Shahin and Liden, Lars and Gao, Jianfeng},
  journal={Transactions of the Association for Computational Linguistics},
  volume={9},
  pages={807--824},
  year={2021},
  publisher={MIT Press One Rogers Street, Cambridge, MA 02142-1209, USA journals-info~…}
}

@article{zhao2022description,
  title={Description-driven task-oriented dialog modeling},
  author={Zhao, Jeffrey and Gupta, Raghav and Cao, Yuan and Yu, Dian and Wang, Mingqiu and Lee, Harrison and Rastogi, Abhinav and Shafran, Izhak and Wu, Yonghui},
  journal={arXiv preprint arXiv:2201.08904},
  year={2022}
}

@article{hudevcek2023llms,
  title={Are LLMs all you need for task-oriented dialogue?},
  author={Hude{\v{c}}ek, Vojt{\v{e}}ch and Du{\v{s}}ek, Ond{\v{r}}ej},
  journal={arXiv preprint arXiv:2304.06556},
  year={2023}
}

@article{hosseini2020simple,
  title={A simple language model for task-oriented dialogue},
  author={Hosseini-Asl, Ehsan and McCann, Bryan and Wu, Chien-Sheng and Yavuz, Semih and Socher, Richard},
  journal={Advances in Neural Information Processing Systems},
  volume={33},
  pages={20179--20191},
  year={2020}
}

@article{gupta2022instructdial,
  title={InstructDial: Improving zero and few-shot generalization in dialogue through instruction tuning},
  author={Gupta, Prakhar and Jiao, Cathy and Yeh, Yi-Ting and Mehri, Shikib and Eskenazi, Maxine and Bigham, Jeffrey P},
  journal={arXiv preprint arXiv:2205.12673},
  year={2022}
}

@article{luo2024duetsim,
  title={DuetSim: Building User Simulator with Dual Large Language Models for Task-Oriented Dialogues},
  author={Luo, Xiang and Tang, Zhiwen and Wang, Jin and Zhang, Xuejie},
  journal={arXiv preprint arXiv:2405.13028},
  year={2024}
}
